%% file: main-paper.tex
\documentclass[sigconf,natbib=true]{acmart}

\usepackage{booktabs} % For formal tables
\usepackage{amsmath}
\usepackage{graphics}
\usepackage{epsfig}
\usepackage{graphicx}
\usepackage[ruled,vlined, linesnumbered]{algorithm2e}
\usepackage{xcolor}
\usepackage{subfigure}
\usepackage{balance}
\usepackage{multirow}
\usepackage{mathrsfs}
\usepackage{acronym}
\usepackage{placeins}
\usepackage{tabularx}
\usepackage{makecell}
\usepackage{xcolor}
\usepackage[inline]{enumitem}
\usepackage{soul}
\AtBeginDocument{%
  \providecommand\BibTeX{{%
    \normalfont B\kern-0.5em{\scshape i\kern-0.25em b}\kern-0.8em\TeX}}}

\copyrightyear{2024}
\acmYear{2024}
\setcopyright{acmlicensed}
\acmConference[CIKM '24]{Proceedings of the 33rd ACM International Conference on Information and Knowledge Management}{October 21--25, 2024}{Boise, ID, USA}
\acmBooktitle{Proceedings of the 33rd ACM International Conference on Information and Knowledge Management (CIKM '24), October 21--25, 2024, Boise, ID, USA}
\acmDOI{10.1145/3627673.3679692}
\acmISBN{979-8-4007-0436-9/24/10}

\newcommand{\changed}[1]{\textcolor{black}{#1}}   

\newcommand{\header}[1]{\vspace*{1mm}\noindent\textbf{#1.}}

\begin{document}

\title[Content-Based Collaborative Generation for Recommender Systems]{Content-Based Collaborative Generation for\\ Recommender Systems}

% \author{Anonymous Author(s)}
% \renewcommand{\shortauthors}{Anonymous Author(s)}

\settopmatter{authorsperrow=4}
\author{Yidan Wang}
\affiliation{%
  \institution{Shandong University}
  % \streetaddress{30 Shuangqing Rd}
  \city{Qingdao}
  \country{China}}
\email{yidanwang@mail.sdu.edu.cn}

\author{Zhaochun Ren}
\affiliation{%
  \institution{Leiden University}
  \city{Leiden}
  \country{Netherlands}}
\email{z.ren@liacs.leidenuniv.nl}

\author{Weiwei Sun}
\affiliation{%
  \institution{Shandong University}
  % \streetaddress{30 Shuangqing Rd}
  \city{Qingdao}
  \country{China}}
\email{sunnweiwei@gmail.com}

\author{Jiyuan Yang}
\affiliation{%
  \institution{Shandong University}
  % \streetaddress{30 Shuangqing Rd}
  \city{Qingdao}
  % \state{Shandong}
  \country{China}}
\email{jiyuan.yang@mail.sdu.edu.cn}

\author{Zhixiang Liang}
\affiliation{%
  \institution{Zhejiang University}
  % \streetaddress{30 Shuangqing Rd}
  \city{Hangzhou}
  % \state{Zhejiang}
  \country{China}}
\email{zliang18@illinois.edu}

\author{Xin Chen}
\affiliation{%
  \institution{WeChat, Tencent}
  % \streetaddress{30 Shuangqing Rd}
  \city{Beijing}
  \country{China}}
\email{andrewxchen@tencent.com}

\author{Ruobing Xie}
\affiliation{%
  \institution{Tencent}
  % \streetaddress{30 Shuangqing Rd}
  \city{Beijing}
  \country{China}}
\email{xrbsnowing@163.com}

\author{Su Yan}
\affiliation{%
  \institution{WeChat, Tencent}
  % \streetaddress{30 Shuangqing Rd}
  \city{Beijing}
  \country{China}}
\email{suyan@tencent.com}

\author{Xu Zhang}
\affiliation{%
  \institution{WeChat, Tencent}
  % \streetaddress{30 Shuangqing Rd}
  \city{Beijing}
  \country{China}}
\email{xuonezhang@tencent.com}

\author{Pengjie Ren}
\affiliation{%
  \institution{Shandong University}
  % \streetaddress{30 Shuangqing Rd}
  \city{Qingdao}
  \country{China}}
\email{renpengjie@sdu.edu.cn}

\author{Zhumin Chen}
\affiliation{%
  \institution{Shandong University}
  % \streetaddress{30 Shuangqing Rd}
  \city{Qingdao}
  \country{China}}
\email{chenzhumin@sdu.edu.cn}

\author{Xin Xin}
\authornote{The corresponding author.}
\affiliation{%
  \institution{Shandong University}
  % \streetaddress{30 Shuangqing Rd}
  \city{Qingdao}
  \country{China}}
\email{xinxin@sdu.edu.cn}

\renewcommand{\shortauthors}{Yidan Wang et al.}

\begin{abstract}
Generative models have emerged as a promising utility to enhance recommender systems.
%This task has been formulated as a sequence-to-sequence generation process, wherein the input sequence comprises data pertaining to the user's previously interacted items, and the output sequence represents the generative identifier for the suggested item. 
% \todo{However, existing generative recommendation methods still fail to: 
% \begin{enumerate*}[label=(\roman*)]
% \item effectively model user-item collaborative signals and item content information in a unified generative framework;
% \item perform effective alignment between content information and collaborative signals.
% \end{enumerate*} }
It is essential to model both item content and user-item collaborative interactions in a unified generative framework for better recommendation.
Although some existing large language model (LLM)-based methods contribute to fusing content information and collaborative signals, 
they fundamentally rely on textual language generation, which is not fully aligned with the recommendation task. 
How to integrate content knowledge and collaborative interaction signals in a generative framework tailored for item recommendation is still an open research challenge.

% Although there are some LLM-based recommendation models trying to combine content information and collaborative signals, these approaches 

In this paper, we propose \underline{\textbf{co}}ntent-based col\underline{\textbf{la}}-borative generation for \underline{\textbf{rec}}ommender systems, namely ColaRec.
ColaRec is a sequence-to-sequence framework which is tailored for directly generating the recommended item identifier. Precisely, the input sequence comprises data pertaining to the user's interacted items, and the output sequence represents the generative identifier (GID) for the suggested item. 
To model collaborative signals, the GIDs are constructed from a pretrained collaborative filtering model, and the user is represented as the content aggregation of interacted items. 
% Then the aggregated item textual description is fed to a  model to capture the content information.
To this end, ColaRec captures both collaborative signals and content information in a unified framework. 
Then an item indexing task is proposed to conduct the alignment between the content-based semantic space and the interaction-based collaborative space.
Besides, a contrastive loss is further introduced to ensure that items with similar collaborative GIDs have similar content representations.
%, achieving better alignment.  
% and a contrastive loss are proposed to conduct the mapping between the content-based
% semantic space and the interaction-based collaborative space.
To verify the effectiveness of ColaRec, we conduct experiments on four benchmark datasets. Empirical results demonstrate the superior performance of ColaRec. 
% Our code used in this work is available at \url{https://github.com/nancheng58/DebiasedSR_DRO}.
%Addtionally, we find that \todo{xxx}
\end{abstract}

% \ccsdesc[500]{Information systems~Retrieval models and ranking}
% \ccsdesc[500]{Information systems~Novelty in information retrieval}


\maketitle
\input{sections/introduction}

\input{sections/related_work}

\input{sections/method}

\input{sections/experiment}

\input{sections/conclusion}

\input{sections/acknowledge}

\bibliographystyle{ACM-Reference-Format}
\balance
\bibliography{refer-clear}

\end{document}

%% file: sections/introduction.tex
\section{introduction}
\begin{figure}
  \centering
  \includegraphics[width=\linewidth]{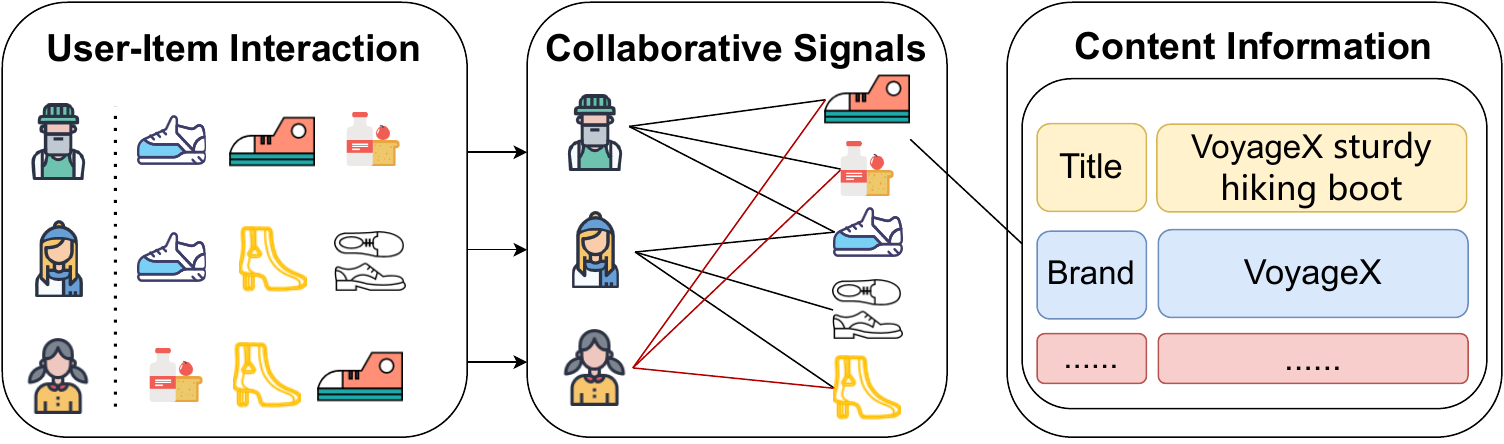}
  \caption{{Illustration of collaborative signals and content information. Collaborative signals refer to the knowledge contained in user-item interactions while content information refers to the textual description of items.}}
  % \Description{A woman and a girl in white dresses sit in an open car.}
  \label{fig:intro}
  \vspace{-0.3cm}
\end{figure}

Recommender systems are widely deployed to 
discover user interests and provide personalized information services~\cite{reinforce-e-commerce,qi2020fednews,nextitnet}.
Recently, generative models, such as large language models (LLMs)~\cite{zhao2023survey,chang2023survey} and diffusion models~\cite{yang2023diffusion,cao2022survey}, 
have gained prominence in advancing artificial intelligence. 
In such a context, plenty of research has emerged to utilize generative models for recommendation~\citep{wang2023diffrec,rajput2023tiger}.
To achieve satisfying results, the recommender should be able to incorporate both user-item collaborative signals and item content information. Collaborative signals refer to the knowledge contained in the user-item interactions while item content information refers to the textual description of items, as shown in Figure~\ref{fig:intro}.

LLM-based recommendation is a straightforward solution to fuse  content information with collaborative signals. 
In general, these approaches reorganize the historical user-item interactions into text-based natural language, aka the instruction-tuning dataset. Then, the recommendation task is reformulated as language generation under designed prompts ~\cite{fan2023llmrec-survey,li2024genrecsurvey1}. Although this kind of approach utilizes the powerful text generation capability of LLMs, there exists the inherent misalignment between the two tasks of language generation and item recommendation.
For example, there is usually a non-trivial grounding stage to map the generated language to a concrete item \cite{lin2023multi,bao2023biground}. LLMs also suffer poor ranking performance to generate target itemIDs from a large candidate pool {\cite{zhang2023InstructRec,liao2023llara}}. The task discrepancy limits the practical usage of LLM-based methods. 

\begin{figure}
%\vspace{0.4cm}
  \centering
  \includegraphics[width=0.8\linewidth]{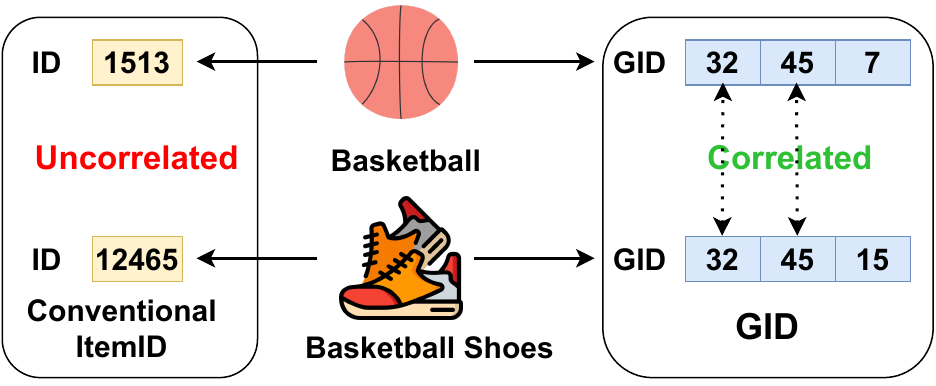}
  \caption{{Comparison between conventional itemIDs and GIDs. 
  %Unlike conventional item IDS,
  GIDs contain more concrete correlations.}}
  % \Description{A woman and a girl in white dresses sit in an open car.}
  \label{fig:intro_gid}
  \vspace{-0.3cm}
\end{figure}

Inspired by generative retrieval~\citep{tay2022transformer}, 
some works \cite{rajput2023tiger,si2023SEATER} assign each item with a unique sequence of tokens as the item's generative identifier, aka GID, then a generative sequence-to-sequence model which is tailored for directly generating item recommendation is trained without the need of explicit language modeling. The input sequence comprises historical user-item interactions while the output sequence refers to the GID of the recommended item. In this paper, we refer such the above paradigm as \textbf{\emph{Generative Recommendation}}\footnote{\citet{wang2023AIGC} proposed another paradigm to directly generate new content, e.g., images, for recommendation. 
However, their methods are tailored for the generation of virtual content  and cannot be used for  concrete items. In this paper, we target on the recommendation of concrete items.}.
Compared with conventional itemIDs with an assigned single random token, 
the sequential tokens of a GID contain more explicit representation information.
As shown in Figure \ref{fig:intro_gid}, we see correlated GIDs denote correlated items, and thus help the recommender to conduct more effective generation.
Generative recommendation provides an end-to-end paradigm to design generative models  tailored for the recommendation task. 
Plenty of studies  have been conducted along this research line ~\citep{rajput2023tiger,si2023SEATER,Aleksandr2023gpt4rec,tan2024idgen,zheng2023lcrec}.

However, existing generative recommendation methods 
still fail to effectively model collaborative signals and content information in a unified framework. For example, several methods utilize either the item titles or hierarchical content embeddings of items to construct GIDs~\cite{li2023gpt4rec,rajput2023tiger}. These methods only consider the item content information while the collaborative signals between users and items are overlooked. 
On the contrary, \citet{si2023SEATER} constructed GIDs using item embeddings of a pretrained SASRec~\citep{wang2018sasrec}. Although the GID in this work contains the item-item sequential (collaborative) connections, the proposed recommendation framework in their research fails to model the item content information.

% fail to effectively model and align collaborative signals and item content information in a unified framework.

Besides, existing generative recommendation methods cannot effectively align item content information and collaborative signals. Although \citet{hua2023howindex} proposed a combined approach that concatenates content-based semantic strings with collaborative IDs derived from the item-item co-occurrence matrix. However, this naive concatenation, lacking a proper learning process, fails to achieve effective alignment, resulting in sub-optimal performance.
The alignment should be achieved through an explicit learning process, e.g., the mapping between the content-based semantic space and the interaction-based collaborative space.

In this paper, we propose \underline{\textbf{co}}ntent-based col\underline{\textbf{la}}-borative generation for \underline{\textbf{rec}}ommender systems (ColaRec), a model that unifies both item content information and user-item collaborative signals in a sequence-to-sequence generation framework tailored for the recommendation task.
Taking a particular user as an example, the input sequence of ColaRec consists of 
unordered\footnote{In this paper, we focus on the general recommendation task other than sequential recommendation. To this end, we use unordered item tuples as the input.} 
tuples with each tuple describing the content information of one interacted item of this user. 
Then an encoder-decoder Transformer~\citep{vaswani2017attention} is used to generate the GID of the target item.
%
% \todo{To address the first limitation,}
More precisely, 
we propose two tailored designs to jointly model user-item collaborative signals and item content information. 
Firstly, we propose to construct GIDs using item representations obtained from a pretrained collaborative filtering model. In this paper, we use LightGCN \cite{he2020lightgcn} as the pretrained model. The LightGCN model is trained on the user-item interaction graph and thus the constructed GID can effectively encode the user-item collaborative signals. Note that the LightGCN model can also be alternated with other models. Secondly, in ColaRec, the user is represented as the content aggregation of tuples with each tuple describing one her/his historically interacted item. This design also keeps inline with the nature of collaborative filtering. The aggregation of content-based tuples is fed to the generative Transformer model to effectively capture the item content information.

To conduct the alignment between content information and collaborative signals, we propose an auxiliary item indexing task which targets on mapping the item side information into the GID of this item through the same encoder-decoder model. 
Specifically, the item side information contains both the textual content information and a set of users who have interacted with this item. 
To this end, the indexing task maps both the item content information and user-item interaction signals into the constructed GID, achieving better alignment.
Besides, we further propose a contrastive loss to ensure that items with similar collaborative GIDs are also similar in the content-based semantic space. 

To demonstrate the effectiveness of the proposed ColaRec, we conduct extensive experiments on four public datasets. Experimental results show that the proposed ColaRec outperforms related state-of-the-art baselines\footnote{The code of this work is available at \url{https://github.com/Junewang0614/ColaRec}}. 

% \header{Main Contributions}
Our main contributions are as follows:
\begin{itemize}[leftmargin=*,nosep]
    \item  We propose ColaRec, a generative recommendation framework which utilizes an encoder-decoder model to jointly capture content information and  collaborative signals for recommendation.

    \item We propose an auxiliary item indexing task and 
    a contrastive loss to perform better alignment between item content information and user-item collaborative signals to further enhance the performance of generative recommendation.

    \item We conduct extensive experiments on four datasets to demonstrate the effectiveness of the proposed ColaRec.
    Experimental results show superior recommendation performance of  ColaRec.
\end{itemize}

%% file: sections/related_work.tex
\section{related work}
In this section, we review related literature on collaborative filtering and  generative models for recommendation.
\subsection{Collaborative Filtering}
Collaborative filtering (CF) is one of the most representative methods to build a recommendation agent.
CF believes that a user can be represented as the aggregation of her/his interacted items, and vice versa. The keystone to conduct collaborative filtering is the user-item interaction matrix. Early approaches \cite{koren2009mf,rendle2012bpr} are based on matrix factorization (MF) to jointly model the latent space for users and items. 
Due to the expressiveness of deep neural networks, plenty of research~\citep{he2017neumf,mao2021simplex,Zheng2017DeepCoNN,Chen2018narre,Gao2020ssg}  has been conducted to enhance CF through deep learning.
Besides, since the user-item interaction signals can be naturally encoded into an interaction graph, graph neural networks (GNN) also shed lights in the field of CF. 
\citet{berg2017gcmc} proposed to use graph convolution for matrix completion. \citet{wang2019ngcf} proposed the NGCF model to use GNN for CF. 
\citet{he2020lightgcn} proposed the LightGCN model to simplify GNN for recommender systems, leading to a simpler and linear CF model.  LightGCN serves as one of the most popular GNN-based CF approaches due to its effectiveness. 
\citet{lin2022ncl} proposed NCL to introduce contrastive learning into graph-based CF. 
% CF with content 
Besides, content information has also been utilized to enhance the CF model. \citet{rendle2010FM} proposed the notable factorization machine (FM) to extend MF for categorical contextual features. 
\citet{he2016vbpr,Wei2019mmgcn,Wei2020grcn, Wu2017rrn}  proposed to enhance CF with visual or text features. \citet{li2023recformer} proposed RecFormer to model long text sequences for recommendation.

Different from existing CF methods, in this paper we focus on the new paradigm of generative recommendation, where the item is represented as a sequence of tokens, i.e., GID, and the recommendation is provided in a generative fashion.

% \subsection{Generative Models}
\subsection{Generative Models for Recommendation}
% 1. 传统方法 vaes + gans + diffusion 精简组合
Generative models have become a hot research topic to generate new content, such as images and text.
Variational Autoencoders (VAEs)~\citep{Kingma2014vae,higgins2016betavae} and Generative Adversarial Networks (GANs)~\cite{Goodfellow2014gan,karras2019styleGAN,creswell2018gansurvey}, are two kinds of representative generative models. 
Besides, diffusion models~\citep{Ho2020ddpm,yang2023diffusion,cao2022survey} have also shown promising results in content generation. 
These generative models have also been utilized for recommender systems ~\citep{Liang2018multivae,Shenbin2020recvae,Cai2022pevae,He2018amf,Guo2022IPGAN,Wang2022dasp,wang2023diffrec,li2023diffurec}.

Recently, Transformers~\citep{vaswani2017attention} have shown promise in language generation, leading to notable LLMs like GPTs (Generative Pre-trained Transformers). LLM-based recommendation provides a straightforward
solution to fuse user-item interaction signals with item textual content ~\citep{liao2023llara,zheng2023lcrec,zhang2023collm,luo2024integrating,zhu2023CLLM4Rec}. 
Typically, these approaches need to construct the recommendation training data with natural language, i.e., the instruction-tuning dataset, then the recommendation task is transformed into text generation under designed language prompts ~\citep{zhang2023InstructRec, lin2023multi,fan2023llmrec-survey,li2024genrecsurvey1}. 
For instance, \citet{luo2024integrating} proposed to inject collaborative signals into LLM-based recommendation through prompt augmentation.
\cite{zhang2023collm} and \cite{liao2023llara} proposed to fuse item embeddings into  prompt embeddings for CTR prediction and recommendation. 
\citet{lin2023multi} utilized LLMs to generate multiple aspects of the recommended items including titles and attributes.
LC-Rec~\cite{zheng2023lcrec} proposed to perform recommendation through conducting various language generation tasks.
Despite the strong generation capabilities of LLMs, there is an inherent misalignment between language generation and item recommendation, which limits the practical usage of LLM-based methods. For example, it is usually necessary to go through a non-trivial grounding stage to map the generated language into specific items~\citep{bao2023biground,lin2023multi,liao2023llara}. Besides, 
it is also challenging for LLMs to directly generate recommended items from a large candidate pool~\citep{liao2023llara,zhang2023InstructRec,luo2024integrating}.

% 3. 类DSI 的生成式推荐
Inspired by generative retrieval, generative recommendation is a new paradigm to design generative models which is tailored for the recommendation task. 
% 生成式检索
Generative retrieval, which relies on the parametric memory of generative models to directly generate relevant document identifiers, has drawn increasing attention.
\citet{tay2022transformer} firstly proposed a differentiable search index (DSI) for generative retrieval.
Plenty of methods have been conducted following this research line. Representative works include SEAL~\citep{bevilacqua2022SEAL}, NCI~\citep{wang2022nci}, and GenRet~\citep{Sun2023genret}.
% 生成式推荐
Similar to generative retrieval, generative recommendation constructs a sequence of tokens, aka generative identifier (GID), for each item, then a generative sequence-to-sequence model which is tailored for directly generating item recommendation is trained without the need of explicit language modeling. The input comprises historical user-item interactions and the output refers to the GID of the recommended item.
Following this paradigm, 
TIGER~\cite{rajput2023tiger} is a representative generative recommendation method which uses an RQ-VAE~\citep{Neil2022rqvae} to construct the GID and then uses a encoder-decoder based transformer to generate sequential recommendation.
\citet{si2023SEATER} proposed to construct the GID from a pretrained SASRec \cite{wang2018sasrec} model for sequential recommendation. \citet{hua2023howindex} further investigated the effect of item identifier construction. \citet{tan2024idgen} utilized human language tokens to represent each item with a unique textual identifier.

% 总结
Despite the rise of generative recommendation, existing generative recommendation methods still suffer from the ineffective infusion of collaborative signals and item content information. 
How to jointly model and align collaborative signals and content information in an end-to-end generative framework tailored for item recommendation is still an open research challenge.

\begin{figure*}[t]
  \centering
  \includegraphics[width=1\linewidth]{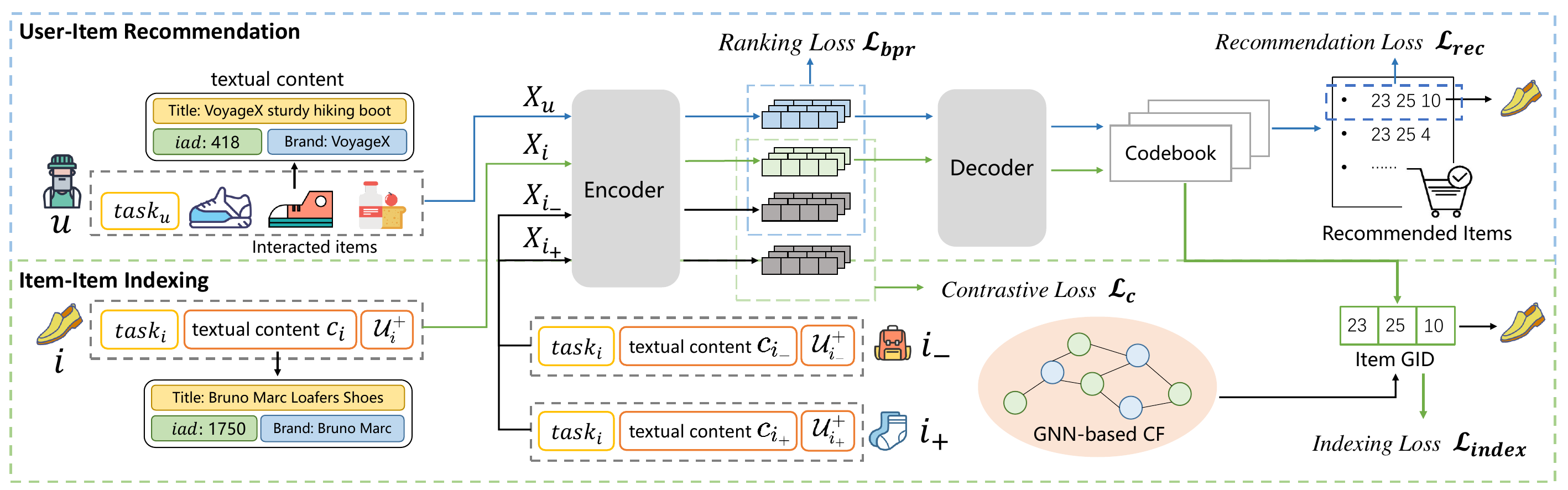}
  \caption{Overview of ColaRec. ColaRec assigns each item with a GID obtained from a GNN-based CF model. ColaRec consists two tasks. User-Item Recommendation aims to map the user’s interacted items with textual content into the GID of the recommended item, i.e., $\mathcal{L}_{\text{rec}}$. Item-Item Indexing targets on the mapping from item side information into the item’s GID, i.e., $\mathcal{L}_{\text{index}}$. %Both the two tasks are achieved through a fine-tuned language model. 
  Besides, a ranking loss $\mathcal{L}_{\text{bpr}}$ and a contrastive loss $\mathcal{L}_{\text{c}}$ are also introduced.} 
  % \Description{A woman and a girl in white dresses sit in an open car.}
  \label{fig:model-framework}
\end{figure*}

%% file: sections/method.tex
\section{Notations and Task Formulation}
 % We first introduce the notations. Then, we describe the task definition of generative recommendation in this paper.

\subsection{Notations} 
% Let $U$ denotes the user set and $I$ denotes the item set. 
Let $u$ and $i$ denote a specific user and an item, respectively. 
The set of interacted items of user $u$ is denoted as $\mathcal{I}_u^+$, and the set of users who have interacted with item $i$ is denoted as $\mathcal{U}_i^+$.
The content description of item $i$ is denoted as $c_i$. The randomly assigned single token to denote user $u$ is the user's atomic identifier, aka, $uad_u$. 
Similarly, the randomly assigned single token to denote item $i$ is the item's atom identifier $iad_i$.
Besides $iad_i$, each item $i$ is also assigned with a generative identifier GID$_i=[z_i^1,z_i^2,\cdots,z_i^l]$, where $l$ denotes the length of GID$_i$.

\subsection{Generative Recommendation}
The task of generative recommendation is given the 
input describing the information of $\mathcal{I}_u^+$, generating a list of GIDs as the recommendation result. 
The GID is generated through an auto-regressive manner. 
The probability of recommending item $i$ for user $u$ is estimated as: 
{\begin{equation}
p(u,i)= \prod_{t=1}^{l}p(z_i^t|\mathcal{I}_u^+, z_i^1,z_i^2,\cdots,z_i^{t-1}).
\end{equation}}
The recommender selects items with the top-$n$
highest $p(u,i)$ as the recommendation list for user $u$.

\section{Methodology}
In this section, we describe the details of ColaRec. We first provide the overview of the proposed ColaRec. Then the construction of items' GIDs is detailed. After that, the user-item recommendation task and the item-item indexing task are described.
% \changed{including a tailored constrastive loss for better alignment between collaborative signals and content information.} 
Finally, we describe the joint optimization of the above tasks.

\subsection{Overview of ColaRec}

Figure \ref{fig:model-framework} illustrates an overview of the proposed ColaRec. 
ColaRec constructs the GID using a graph-based CF model, which effectively captures collaborative signals.
% thus the introduced GID can effectively capture 
% collaborative signals.
The training of ColaRec consists of two tasks: the user-item recommendation task and the item-item indexing task. 
The user-item recommendation aims to map the content information of the user's historical interacted items into the GID of the recommended item. 
The item-item indexing task targets on the mapping from the item side information into the item's GID. 
Both the two tasks are achieved through a shared encoder-decoder based model.
The recommendation task unifies both collaborative signals and item content information for better recommendation, 
while the indexing task performs the alignment between collaborative signals and content information. 
%\todo{Note that parameters of the involved language model are also fine-tuned to better adapt the language model for recommendation.}

\subsection{Generative Identifier Construction}\label{gid-construction}
The construction of GIDs plays a crucial role in generative recommendation.
Generally speaking, GIDs should satisfy the following expectations for better recommendation:
\begin{enumerate*}[label=(\roman*)]
\item GIDs need to contain knowledge about 
both collaborative signals and content information;
\item Correlated items (e.g., similar items in content or items interacted by similar users) should have correlated GIDs; 
\item Each item should have one unique GID and each GID should correspond to one specific item. 
\end{enumerate*}

To fulfill above expectations, we utilize a hierarchical clustering approach to construct GIDs from a graph-based CF model. 
Specifically, we first extract item representations from a pretrained LightGCN model.
% Then the constrained $K$-means algorithm is employed to iteratively cluster the items into 
Then the constrained $K$-means algorithm is called hierarchically based on the item representations. 
The next level clustering is conducted with items in the current cluster as the whole item set.
For the $t$-th level clustering with $t\in[1,l-1]$, the number of items is no more than $K^{l-t}$ \changed{in each cluster}.
Regarding the last level of leaf nodes, we randomly allocate 1 to $K$ to the items.
In this way, we establish a $K$-ary tree to organize the item set.
Each item corresponds to a leaf node, while the path from the root to the leaf node is the GID of the item. 
Since LightGCN is trained on the user-item interaction graph, the GID can naturally encode collaborative signals.
Meanwhile, 
%in order to combine content features of items, 
there is a codebook embedding matrix for every position of the GID, which will incorporate the content information in the item indexing task.
We give the detailed description in section \ref{subsec:indexing}. 
To this end, the GID together with the corresponding codebook embeddings helps the recommender to model both collaborative signals and content information.

% model inputs 分为了三个层次，一层一层一直到user层次
\subsection{User-Item Recommendation}
\header{Model Inputs}
The input sequence of user-item recommendation for user $u$ consists of unordered tuples with each tuple describing the content information of one interacted item of this user. 
For the textual description of item $i$, we adopt the universal data format from \cite{li2023recformer}.
Specifically, textual description $c_i$ of item $i$ is formulated as a sequence that comes from a flatted attribute dictionary consisting of key-value attribute pairs $(k,v)$, i.e., $[k_1{:}v_1,k_2{:}v_2, \dots]$. 
Besides, we also introduce the item atomic identifier $iad_i$ into the content information to further increase model fidelity.
To this end, the content tuple of item $i$ is formulated as:
\begin{equation}
    c_i = [iad_i,k_1{:}v_1,k_2{:}v_2, \dots].
\end{equation}
The key idea of CF is that users' preferences can be inferred from their interacted items.
Therefore, for each user $u$, the input consists of the content aggregation of item tuples that $u$ has interacted with to reinforce the collaborative signals.

%3. 为了区分任务，输入前加入user的prompt
Since the training of ColaRec has two tasks, in the user-item recommendation task, we augment the input with a special task token $task_u$  at the beginning of the input to inform the model that the ongoing task is the recommendation task.
Thus, the input for the user-item recommendation task is:
\begin{equation}
    X_u = [task_{u},\{c_i|i \in \mathcal{I}_u^+\}].
\end{equation}

\header{Item Generation}
We employ an encoder-decoder based Transformer model to generate item recommendation.
Given the model input $X_u$, the model encoder captures the semantic information of $X_u$ and returns the hidden state $\text{Encoder}(X_u)$.
%, while the decoder generates a sequence of tokens as the GID.
After that, given the generated tokens $z^{<t}$ before the $t$-th generation step, the decoder generates the latent representation $\mathbf{d}_t \in \mathbb{R}^m$ for the $t$-th token of GID.  $m$ is the dimension of the latent representation. This process can be formulated as:
\begin{equation}
\label{eq:d_t}
    \mathbf{d}_t = \text{Decoder}(\text{Encoder}(X_u),z^{<t})
\end{equation}
The generation probability at step $t$ is estimated by $\mathbf{d}_t$ and the codebook embedding matrix for position $t$, which is formulated as:
\begin{equation}
\label{eq:generate_p}
    p(z^t|z^{<t},X_u) = \operatorname{softmax}(\mathbf{d}_t\cdot \mathbf{E}_{t}^\top ),
\end{equation}
where $\mathbf{E}_{t}$ is the $t$-th step codebook embedding matrix.

We adopt the cross-entropy loss for model optimization. Specifically, given a $(u,i)$ pair in the training set, the generative loss of recommendation is formulated as:
\begin{equation}
\mathcal{L}_{\text{rec}} = -\sum_{t=1}^{l}\log p(z^t_i|X_u,z^1_i,z^2_i,\cdots,z^{t-1}_i).
\end{equation}

In this work, we use a pretrained T5 \cite{Raffel2020t5} as the Transformer model. 
The parameters of T5 are also fine-tuned through back propagation to better adapt the model for recommendation.
Note that our approach is model-agnostic and T5 can be replaced by other sequence-to-sequence models, even without pretraining.

\subsection{Item-Item Indexing}
\label{subsec:indexing}
To align collaborative signals and item content information, we introduce an item-item indexing task which conducts the mapping from the content-based semantic space into the interaction-based collaborative space. 

\header{Model Inputs}
The input sequence for item indexing contains the textual information of the item. Besides, we also introduce information of users who have interacted with this item, to further encode collaborative signals. 
Similar to the recommendation task, we augment the input with a special task token $task_i$ at the beginning of the input for item indexing.
Therefore, the input of item indexing is formulated as:
\begin{equation}
    X_i = [task_i,c_i,\{uad_u|u\in \mathcal{U}_i^+\}].
\end{equation}

\header{Item Indexing}
The indexing task is conducted through the same  model and codebook embeddings as the recommendation task. The generation probability for the indexing task is formulated similarly with Eq.~(\ref{eq:d_t}) and Eq.~(\ref{eq:generate_p}) except that the model input is $X_i$ instead of $X_u$.
% other than $X_u$.
We adopt the cross-entropy loss for parameter learning. 
The loss for item indexing is defined as:
\begin{equation}
\mathcal{L}_{\text{index}} = -\sum_{t=1}^{l}\log p(z^t_i|X_i,z^1_i,z^2_i,\cdots,z^{t-1}_i).
\end{equation}

\subsection{Multi-Task Training}
Besides the above two tasks, we further introduce a ranking loss to enhance the ranking ability of ColaRec and 
a contrastive loss to conduct better alignment.
% \changed{Specifically, for a $(u, i)$ pair in the training dataset, we sample an item $i_-$ from the contrastive loss (i.e., Eq.~(\ref{eq:contrastive})) that the user $u$ has not interacted with as the negative sample.}

\header{Item Ranking}
For a $(u, i)$ pair in the training dataset, we randomly sample one item that the user $u$ has not interacted with 
% and has no overlap with the positive item $i$, 
% , and for which the GID has no overlap with the positive item $i$, 
as the negative sample $i_-$.
% Specifically, we use the sampled item $i_-$ in
% the contrastive loss Eq.~(\ref{eq:contrastive}) as the negative sample. 
%
The BPR loss~\cite{rendle2012bpr} is utilized to optimize the item ranking, which is formulated as:
\begin{equation}
\mathcal{L}_{\text{bpr}} = -\ln \sigma (\mathbf{h}(X_u) \cdot (\mathbf{h}(X_i) - \mathbf{h}(X_{i_-}))),
\end{equation}
where $\mathbf{h}(\cdot)$ denotes the last  hidden states of Encoder($\cdot$), and $\sigma$ denotes the sigmoid function. 
The above loss pushes together the positive $(u,i)$ pair  and pushes away the negative $(u,i_-)$ pair, thus helping the model to capture the ranking knowledge between items.

\header{Contrastive Learning}
To conduct better alignment between collaborative signals and content information, a contrastive loss is further introduced.
The idea is that items with similar GIDs should also be similar in the content-based semantic space. 
To this end, for the item $i$, we randomly sample an item $i_+$ which has the overlapped prefix tokens in GIDs as the positive sample. The sampled item $i_-$ in $\mathcal{L}_\text{bpr}$ serves as the negative sample. Note that here we ensure the sampled item $i_-$ in $\mathcal{L}_\text{bpr}$ has no overlapped GID tokens with item $i$.
The contrastive loss is defined as:
\begin{equation}
\label{eq:contrastive}
\mathcal{L}_{\text{c}} = - \ln \sigma (\textbf{h}(X_i)\cdot (\textbf{h}(X_{i_+}) - \textbf{h}(X_{i_-}))),
\end{equation}
Such a contrastive loss helps the model to learn better item input representations.

\header{Joint Optimization}
Finally, ColaRec is trained with the above described tasks jointly:
\begin{equation}
\mathcal{L} = \mathcal{L}_{\text{rec}} + \mathcal{L}_{\text{index}} + \mathcal{L}_{\text{bpr}} + \alpha \mathcal{L}_{\text{c}},
\end{equation}
where $\alpha$ denotes the weight for the contrastive loss.

During the inference, to avoid the recommender from generating invalid GIDs, we employ the constrained beam search~\citep{cao2021gener} to limit the generative range of the current token based on the prefix tokens.

%% file: sections/experiment.tex
% \begin{table*}
%   \caption{Some Typical Commands}
%   \label{tab:commands}
%   \begin{tabular}{ccl}
%     \toprule
%     Command &A Number & Comments\\
%     \midrule
%     \texttt{{\char'134}author} & 100& Author \\
%     \texttt{{\char'134}table}& 300 & For tables\\
%     \texttt{{\char'134}table*}& 400& For wider tables\\
%     \bottomrule
%   \end{tabular}
% \end{table*}

\begin{table}
\setlength\tabcolsep{5pt}
\caption{Statistics of four public datasets after preprocessing.}
\label{tab:datasets}
\begin{tabular}{lccc}
\toprule
\textbf{Datasets}  & \textbf{\#Users} & \textbf{\#Items} & \textbf{\#Interactions} \\
\midrule
Beauty & 22,363   & 12,101   & 198,502     \\
Sports & 35,598   & 18,357   & 296,337     \\
Phone  & 27,879   & 10,429   & 194,439     \\
Recipe  & 17,813   & 41,240    & 555,618    \\
\bottomrule
\end{tabular}
\end{table}

\input{table/main-results}

\section{Experiment} \label{sec:exp}
In this section, we conduct experiments to evaluate the proposed ColaRec. We aim to answer the following
research questions:
\begin{enumerate}[leftmargin=*, label=RQ\arabic*]
\item How does the proposed ColaRec perform compared with existing recommendation methods?
\item How does the joint training of multiple tasks affect the performance of ColaRec?
\item How does the design of GIDs affect the recommendation performance?
\end{enumerate}
\subsection{Datasets}

We use four real-world public datasets to evaluate the performance of ColaRec.
Specifically, the experiments are conducted on three subcategories from Amazon Product Reviews\footnote{\url{https://jmcauley.ucsd.edu/data/amazon/}} (``Beauty'', ``Sports and Outdoors'', and ``Cell Phones and Accessories'') and ``Recipe'' from Food.com\footnote{\url{https://www.kaggle.com/datasets/shuyangli94/food-com-recipes-and-user-interactions}}. 
Users and items that have less than five interactions are filtered out.
Table \ref{tab:datasets} shows the statistics of all four datasets. As for content information, we use  ``title'', ``brand'' and ``categories'' from the Amazon item metadata as the textual content information of items. For Recipe, we use ``name'', ``description'', and ``tag'' to describe item content.

\subsection{Evaluation Protocols}
We adopt cross-validation to evaluate the performance of recommenders.
In this paper we focus on general recommendation rather than sequential recommendation. To this end, we randomly split each user’s historical interactions into the training/validation/test set with the ratio of 8:1:1. 
We employ two widely used metrics, recall@$n$ and normalized discount cumulative gain (NDCG@$n$), to evaluate the model performance.
Recall evaluates how many ground-truth items occur in the recommended list, while NDCG further focuses on their rankings in the list. 
%In this paper, we set K as 5,10 and 20. 
Note that in this paper, the candidate item set is the whole item set, other than a small subset with selected items. Each experiment is conducted three times and the average score is reported.

\subsection{Baselines}
We compare ColaRec with several representative related baselines, including both conventional CF-based methods and generative models for recommendation. CF-based baselines include 
%MF~\cite{rendle2012bpr}, 
NeuMF~\cite{he2017neumf}, LightGCN~\cite{he2020lightgcn}, SimpleX~\cite{mao2021simplex}, and NCL \cite{lin2022ncl}.
Generative models for recommendation include MultiVAE~\cite{Liang2018multivae}, DiffRec~\cite{wang2023diffrec}, DSI~\cite{tay2022transformer}, TIGER~\cite{rajput2023tiger} and LC-Rec\cite{zheng2023lcrec}.

\begin{itemize}[leftmargin=*,nosep]
    % \item \textbf{MF}~\citep{rendle2012bpr} is the most common model for CF based on matrix factorization and optimized by BPR loss. 
    \item \textbf{NeuMF}~\citep{he2017neumf} enhances MF with multiple hidden layers to learn non-linear patterns in user-item interactions.  
    \item \textbf{LightGCN}~\citep{he2020lightgcn} simplifies GNN for CF and learns user and item representations via linear neighborhood aggregation. 
    \item\textbf{SimpleX}~\citep{mao2021simplex} is a simple CF model with a cosine-based contrastive loss and negative sampling.
    \item\textbf{NCL} \cite{lin2022ncl} improves LightGCN with contrastive learning. 
    \item\textbf{MultiVAE}~\citep{Liang2018multivae} is an autoencoder-based method, which utilizes VAEs to model the interaction signals.
    \item\textbf{DiffRec}~\citep{wang2023diffrec} is a new proposed recommendation model based on diffusion models. DiffRec learns the user-item interaction knowledge through a reconstruction and denoising manner. 
    \item\textbf{DSI} \cite{tay2022transformer} is a generative document retrieval method.
    To adapt DSI for recommendation, we formulate the input as GIDs of the user's historical interacted items. 
    We use two versions of DSI.
    \textbf{DSI-R} refers to the DSI model with a random string as the GID of the item.
    \textbf{DSI-S} is a DSI model which constructs the item GIDs with a hierarchical $K$-means algorithm based on the item textual content embeddings from a pretrained BERT model.
    \item\textbf{TIGER} \cite{rajput2023tiger} is a generative recommendation method. Specifically, a pretrained Sentence-T5 encoder is used to obtain embeddings of the items' textual content. These embeddings are then quantized using an RQ-VAE to build GIDs. We do not introduce sequential orders to adapt TIGER for general recommendation.
    \item\textbf{LC-Rec} \cite{zheng2023lcrec} follows a similar approach to TIGER~\cite{rajput2023tiger} for item index learning. Besides, it employs various language generation tasks under different prompts to accomplish recommendation. 
    
\end{itemize}
We don't introduce P5-based baselines ~\citep{geng2022p5,hua2023howindex} since these methods require the model input prompt to include candidate items for the general recommendation task. Given the limited input length, these methods cannot perform item ranking among the whole item set. 

\subsection{Implementation Details}
For all datasets, the length of GIDs is set to $l=3$, and the number of clusters in hierarchical $K$-means is set as $K=32$, except for Recipe in which the cluster number is set as 48. 
Each user is represented through the aggregation of randomly sampled interacted item tuples, while each item introduces one randomly sampled user who have interacted with it in the indexing task. 
We use a uniform distribution to sample negative instances for $\mathcal{L}_\text{bpr}$ and $\mathcal{L}_{c}$ to avoid the effect of different negative sampling strategies.
We leave the investigation of negative sampling for generative recommendation as the future work.
To be consistent with the word embeddings of the pretrained T5-small model, the embedding dimensions of $uad$, $iad$ and codebooks in ColaRec are set to 512.
The values of the contrastive loss coefficient, i.e., $\alpha$ are set to \{0.02,0.08,0.1,0.05\} in Beauty, Sports, Phone and Recipe respectively.
We optimize the model using AdamW with 5$e$-4 as the learning rate. The batch size is set to 128. 
For baselines, we carefully search the hyper-parameters except for user and item embedding sizes, which are set to 512 to ensure a fair comparison with ColaRec.

\subsection{Performance Comparison (RQ1)} \label{sec:overall}
To answer RQ1, we conduct a comparative analysis of the proposed ColaRec on both overall users and long-tail users.

% \header{Results on Full Set}
\subsubsection{Comparison on whole users}
Table \ref{tab:main-col} shows the performance comparison of overall users. From these results, we make the following observations.
Firstly,  ColaRec achieves the best recommendation performance on all datasets except for NDCG@5 in Beauty, which achieves comparable scores with the best NCL baseline.
% [Firstly, the proposed ColaRec achieves the best recommendation performance on all datasets in terms of almost all evaluation metrics.]
In particular, ColaRec consistently outperforms previous CF-based and \changed{generative models} in Recall@20, achieving a relative improvement of 3.87\%, 3.54\%, 6.72\% and 3.88\% on Beauty, Sports, Phone and Recipe, respectively.
% which are significantly higher than the state-of-the-art results.
These results demonstrate the effectiveness of ColaRec and its generalization across different domains.

Secondly, compared with DSI-R and DSI-S, which directly adapt generative retrieval methods to the recommendation task, ColaRec achieves a notable 47.89\%, 38.13\%, 80.82\% and 26.11\% relative improvement in terms of Recall@5 on four datasets respectively, demonstrating the effectiveness of the proposed ColaRec. Such results also demonstrate that naively transferring the generative retrieval methods for recommendation cannot achieve satisfying results.
Furthermore, our method consistently outperforms 
existing representative generative methods TIGER and LC-Rec.  For TIGER, the reason may be that it only considers item content information but the collaborative signals are overlooked without an explicit alignment process. 
While for LC-Rec, the recommendation is conducted through 
language modeling and generation under designed prompts, which is not fully aligned with the recommendation task, leading to sub-optimal results.

Lastly, existing generative methods (e.g., DiffRec and TIGER), while exhibiting promising results in some specific scenarios \cite{rajput2023tiger,wang2023diffrec,li2023diffurec}, still underperform the strong CF-based methods (e.g., NCL) in the general recommendation task.  
In contrast, ColaRec achieves competitive results compared to these CF methods on all datasets, demonstrating the potential of 
infusing content information for collaborative generation of recommender systems.

\input{table/Inactive-users}
\subsubsection{Comparison on long-tail users}
We also conduct experiments to verify the recommendation performance of ColaRec on long-tail users with sparse interactions.
In this experiment, the ratio between head users and long-tail users is set as 20\%:80\%.
Table~\ref{tab:inact-user} reports the results of Recall on four datasets. Results of NDCG show similar trends and are omitted due to the reason of space.
We can see that ColaRec significantly outperforms all baselines when generating recommendation for long-tail users.
The reason is that ColaRec models both user-item interactions and item content information. Given the long-tail users with less interaction knowledge, ColaRec gains better performance with the help of the content information.

To conclude, the proposed ColaRec is effective to yield better performance compared with existing baselines. This improvement is more significant on long-tail users.

\input{table/merge}

\subsection{Ablation Study (RQ2)}
In this section, we conduct ablation studies to analyse the effectiveness of each component in ColaRec.
% To answer RQ2, i.e., analyzing the contributions of each technique or component in ColaRec, we conduct an ablation study. 
We implement four ablative variants of ColaRec, including:
\begin{enumerate*}[label=(\arabic*)]
\item \textit{w/o textual content} deletes all textual content information in the model input;
\item \textit{w/o indexing} removes the item-item indexing task;
\item \textit{w/o $\mathcal{L}_{\text{bpr}}$} removes the ranking loss $\mathcal{L}_{\text{bpr}}$; and
\item \textit{w/o $\mathcal{L}_{c}$} removes the contrastive loss $\mathcal{L}_{\text{c}}$.
\end{enumerate*}
% Note

The results are shown in the upper part of Table~\ref{tab:abl-id}.
These results clearly indicate that the removal of any component from our proposed method results in a noticeable decline in overall performance.
From the results of variant (1), we see that removing textual content information leads to the significant performance downgrade.
Specifically, a \(20.99\%\), \(17.65\%\), \(14.63\%\) and \(8.59\%\) drop in Recall@5 is observed across the four datasets, respectively. 
This suggests the importance of incorporating textual content information to enhance the models' understanding of items.
Furthermore, the variant (3), which is trained without \(\mathcal{L}_{\text{bpr}}\), exhibits a notable performance decrease compared to ColaRec.
This highlights the effectiveness of the pairwise ranking objective, which focuses on the relative prioritization of positive items within the generative recommendation paradigm. 
Besides, the results of variant (2) and (4) verify the effectiveness of the proposed item-item indexing task and the contrastive objective \(\mathcal{L}_{c}\).
A consistent performance reduction is observed on the four datasets when either of the two techniques is removed.
This suggests that aligning item content information with user-item collaborative signals 
% is essential to generate better recommendation.
not only facilitates mutual reinforcement but also enables the learning of more comprehensive and effective representations.

In conclusion, each component of ColaRec is essential to improve the recommendation performance.

\input{pics/part-length}
\input{pics/part-cluster}

\subsection{GID Investigation (RQ3)} 
The construction of GIDs plays a crucial role in generative recommendation as it defines the model's search space for generation. 
To answer RQ3, we conduct two analytical experiments about GIDs: (1) an analysis of different GID types (e.g., the single token $iad$, the random assigned GID, and the content-based GID), and (2) an analysis of hyper-parameters when constructing GIDs (e.g., the length and the $K$-means cluster number).

% \header{Comparison of Different GID Types}
\subsubsection{Effect of different GID types}
To further evaluate our GID generation strategy, we conduct an ablation study comparing it with three techniques: \textit{iad}-based GID, $Random$ GID, and $Content$ GID.
Specifically, the \textit{iad}-based GID assigns a unique single token to each item to represent it using the corresponding vector in the item embedding matrix. The \textit{Random} GID assigns a random string to each item as the identifier, without considering any prior knowledge. % content information. 
The \textit{Content} GID represents the GID constructed from item content information. Unlike the collaborative GID used in ColaRec, the content-based GID employs a hierarchical $K$-means clustering algorithm~\citep{tay2022transformer} to cluster items based on their textual representation derived from a pretrained BERT model.
To make a fair comparison with ColaRec, the length and the codebook size of both $Random$ GIDs and $Content$ GIDs are identical to those in ColaRec.

The bottom part of Table~\ref{tab:abl-id} details our results on the four datasets, where ColaRec attains the best performance. 
This indicates that our GID construction method, based on collaborative signals, is highly effective for generative recommendation. 
The improvement over the \textit{Content} GID highlights the effectiveness of collaborative signals in recommendation tasks. 
Furthermore, ColaRec's superior performance compared to the \textit{iad} method demonstrates the benefit of explicitly introducing item correlations into item GIDs. 
In addition, the \textit{Random} method leads to the lowest performance as the random string could further introduce noisy signals in the learning process.
These results illustrate the importance of constructing effective GIDs for generative recommenders.

% 2. 长度和K的影响
\subsubsection{Impact of GID Hyper-parameters}
In this section, we investigate the impact of hyper-parameters $l$ and $K$ in the GID construction process.
Figure \ref{fig:length} shows the results of ColaRec with different GID lengths $l$, ranging from $1$ to $4$.
%The model's performance appears robust across different values of $l$.
Note that to include the whole item set, shorter GIDs indicate larger search spaces for each GID position, and vice versa.
We can see that as $l$ varies, the recommendation performance exhibits some fluctuations.
For the performance drop from $l=1$ to $l=2$ in Beauty and Sports, the reason is that compared with single token GIDs, GIDs with $l=2$ increase \changed{one} decoding step but the search space for each step is still large, increasing the search difficulty. 
When $l=3$, ColaRec achieves a better trade-off between decoding steps and the search space for each step, leading to the best performance in most cases.
When $l=4$, a too long GID means more autoregressive decoding steps in model generation, which increases generation difficulty and inference latency.
Therefore, we have selected $l=3$ as the default setting in this paper.

To assess how the number of clusters $K$ affects performance, we fix the GID length $l$ as $l=3$ and vary $K$ with values $32$, $64$, $96$, and 128. As shown in Figure~\ref{fig:center}, a higher $K$ typically results in a slight decrease in overall performance.
The decrease in Beauty is more notable.
The reason is that a higher $K$ indicates a larger search space in the decoding process, and thus increases the generation difficulty. 
To this end, choosing a suitable clustering number according to the number of items is important for generative recommendation. 
The principle is that $K$ needs to be large enough to encode the whole item set, after that $K$ needs to be controlled  to limit the search space.

%% file: table/main-results.tex
\begin{table*}[t]
\caption{Performance comparison on four public datasets. The best and the second-best scores are marked in bold and underlined fonts, respectively. * denotes the 
paired t-test with significance p-value < 0.1. R and N
stand for Recall and NDCG.}
\label{tab:main-col}
\vspace{-0.1cm}
\renewcommand\arraystretch{0.9}
\scalebox{0.97}{
\begin{tabular}{ll | cccc | cc | cccc| c}
\toprule
&  & \multicolumn{4}{c|}{\textbf{CF-based Methods}} & \multicolumn{7}{c}{\textbf{Generative Models for Recommendation}}\\
\textbf{Datasets}             & \textbf{Metric}                             & \textbf{NeuMF}  & \textbf{LightGCN} & \textbf{SimpleX}    & \textbf{NCL}     & \textbf{\textbf{MutiVAE}} & \textbf{DiffRec} & \textbf{DSI-R} & \textbf{DSI-S} & \textbf{TIGER} & \textbf{LC-Rec} & \textbf{Ours}      \\
\midrule
\multirow{6}{*}{Beauty}              & R@5                       & 0.0447 & 0.0649   & 0.0551 &  {\underline{0.0650}}    & 0.0530  &  {0.0524}  & 0.0128 & 0.0451 &  0.0519 & {0.0492} & \textbf{0.0667\rlap{\textsuperscript{*}}}      \\
                     & R@10                      & 0.0653 & \underline{0.0952}   &0.0831 & {0.0940}            & 0.0776  & {0.0741}  & 0.0228 & 0.0705 & 0.0799 &{0.0770} & \textbf{0.0993\rlap{\textsuperscript{*}}}     \\
                     & R@20                      & 0.0889 & 0.1314   & 0.1193& {\underline{0.1320}}     & 0.1093  &  {0.1016}  & 0.0360 & 0.1018 &  {0.1154} & {0.1104} &\textbf{0.1371\rlap{\textsuperscript{*}}}      \\
                     & N@5                         & 0.0315 & \underline{0.0450}   & 0.0377 &  {\textbf{0.0452}}  & 0.0362  &  {0.0378}  & 0.0084 & 0.0305 &  0.0350 & {0.0326} & 0.0449     \\
                     & N@10                        & 0.0383 & \underline{0.0549}   & 0.0469 &  {0.0547}         & 0.0443  &  {0.0450}  & 0.0117 & 0.0385 & 0.0443 & {0.0415} & \textbf{0.0556\rlap{\textsuperscript{*}}}     \\
                     & N@20                       & 0.0445 & 0.0643   & 0.0563 &  {\underline{0.0646}}          & 0.0526  &  {0.0521}  & 0.0151 & 0.0470 & 0.0534 & {0.0499} & \textbf{0.0654\rlap{\textsuperscript{*}}}     \\
\midrule
\multirow{6}{*}{Sports}               & R@5                       & 0.0206 & 0.0418   & 0.0355 &  {\underline{0.0427}}  & 0.0314  &  {0.0273}  & 0.0117 & 0.0320 &  0.0374 & {0.0397} & \textbf{0.0442\rlap{\textsuperscript{*}}}  \\
                     & R@10                      & 0.0321 & 0.0623   & 0.0557 &  {\underline{0.0631}}  & 0.0476  &  {0.0403}  & 0.0178 & 0.0497 & 0.0572 & {0.0617} &\textbf{0.0660\rlap{\textsuperscript{*}}}    \\
                     & R@20                      & 0.0471 & 0.0901   & 0.0836 &  {0.0908}  & 0.0713  &  {0.0569}  & 0.0284 & 0.0766 & 0.0881 & {\underline{0.0931}} &\textbf{0.0964\rlap{\textsuperscript{*}}}    \\
                     & N@5                         & 0.0140  & {0.0288}   & 0.0240 &  {\underline{0.0294}}  & 0.0208  &  {0.0193}  & 0.0079 & 0.0225 &  0.0249 & {0.0264} &\textbf{0.0294}      \\
                     & N@10                        & 0.0177 & 0.0355  & 0.0306 &  {\underline{0.0359}}  & 0.0261  &  {0.0235}  & 0.0099 & 0.0284 &  0.0313 &{0.0335} & \textbf{0.0364\rlap{\textsuperscript{*}}}    \\
                     & N@20                        & 0.0215 & 0.0426   & 0.0377 &  {\underline{0.0431}}  & 0.0321  &  {0.0278}  & 0.0126 & 0.0350 & 0.0392 & {0.0413} &\textbf{0.0442\rlap{\textsuperscript{*}}}    \\      
\midrule
\multirow{6}{*}{Phone}               & R@5                       & 0.0410 & 0.0713   & 0.0643 &  {\underline{0.0717}}  & 0.0569  &  {0.0470}  & 0.0187 & 0.0412 &  {0.0601} & {0.0615} & \textbf{0.0745\rlap{\textsuperscript{*}}}  \\
                     & R@10                      & 0.0603 & \underline{0.1052}   & 0.0976 &  {0.1043}  & 0.0855  &  {0.0668}  & 0.0341 & 0.0625 & 0.0895 & {0.0919} &\textbf{0.1121\rlap{\textsuperscript{*}}}    \\
                     & R@20                      & 0.0871 & \underline{0.1487}   & 0.1420 &  {0.1481}  & 0.1233  &  {0.0928}  & 0.0564 & 0.0966 & 0.1299 & {0.1354} &\textbf{0.1587\rlap{\textsuperscript{*}}}    \\
                     & N@5                         & 0.0282  & 0.0481   & 0.0423 &  {\underline{0.0486}}  &0.0378  &  {0.0315}  & 0.0121 & 0.0282 & 0.0403 & {0.0408} & \textbf{0.0490\rlap{\textsuperscript{*}}}     \\
                     & N@10                        & 0.0344 & 0.0590   & 0.0530 &  {\underline{0.0593}}   & 0.0470  &  {0.0379}  & 0.0170 & 0.0347 & 0.0498 & {0.0506} & \textbf{0.0611\rlap{\textsuperscript{*}}}   \\
                     & N@20                        & 0.0412 & 0.0700   & 0.0643 &  {\underline{0.0704}}  & 0.0566  &  {0.0445}  & 0.0225 & 0.0431  & 0.0600 & {0.0615} &\textbf{0.0729\rlap{\textsuperscript{*}}}   \\

\midrule
\multirow{6}{*}{Recipe}               & R@5                       & 0.0118 & {0.0188}   & 0.0114 &  {\underline{0.0192}}  & 0.0167  &  {0.0142}  & 0.0142 & 0.0157 & 0.0168 & {0.0174} & \textbf{0.0198\rlap{\textsuperscript{*}}}  \\
                     & R@10                      & 0.0210 & {0.0296}   & 0.0202 &  \underline{0.0298}  & 0.0285  &  {0.0235}  & 0.0248 & 0.0270 &  {0.0292} & {0.0289} & \textbf{0.0306\rlap{\textsuperscript{*}}}    \\
                     & R@20                      & 0.0339 & {0.0454}   & 0.0328 &  {0.0459}  & {0.0462}  &  {0.0343}  & 0.0403 & 0.0436 & \underline{0.0464} & {0.0454} & \textbf{0.0482\rlap{\textsuperscript{*}}}    \\
                     & N@5                         & 0.0088  & \underline{0.0149}   & 0.0093 &  {\underline{0.0149}}  & 0.0128  &  {0.0105}  & 0.0107 & 0.0122 & 0.0137 &{0.0138} & \textbf{0.0151\rlap{\textsuperscript{*}}}    \\
                     & N@10                        & 0.0119 & \underline{0.0182}  & 0.0122 &  {\underline{0.0182}}   & 0.0167  &  {0.0135}  & 0.0141 & 0.0158 & 0.0176 & {0.0175} & \textbf{0.0185\rlap{\textsuperscript{*}}}   \\
                     & N@20                        & 0.0154 & 0.0223  & 0.0156 &  \underline{0.0224}  & {0.0214}  &  {0.0165}  & 0.0182 & 0.0202  & 0.0221 & {0.0218} & \textbf{0.0232\rlap{\textsuperscript{*}}}   \\

\bottomrule
\end{tabular}
}
\end{table*}

% 参数table

%% file: table/Inactive-users.tex
\begin{table*}[t]
\vspace{0.1cm}
% \vspace{0.3cm}
\setlength\tabcolsep{5pt}
\caption{Performance comparison of long-tail users. The best and the second-best scores are marked in bold and underlined fonts, respectively. ** denotes the improvements are significant with p-value < 0.05. R stands for Recall.}
\label{tab:inact-user}
\scalebox{0.97}{
\begin{tabular}{ll | cccc | cc | cccc | c}
\toprule
&  & \multicolumn{4}{c|}{\textbf{CF-based Methods}} & \multicolumn{7}{c}{\textbf{Generative Models for Recommendation}}\\
\textbf{Datasets}             & \textbf{Metric}                             & \textbf{NeuMF}  & \textbf{LightGCN} & \textbf{SimpleX}    & \textbf{NCL}     & \textbf{\textbf{MutiVAE}} & \textbf{DiffRec} & \textbf{DSI-R} & \textbf{DSI-S} & \textbf{TIGER} & \textbf{LC-Rec} & \textbf{Ours}      \\
\midrule
\multirow{3}{*}{Beauty}              & R@5       & 0.0416 & 0.0636   & 0.0555 & {\underline{0.0639}}                 & 0.0510  &  {0.0464}  & 0.0131 & 0.0415 & 0.0487 & {0.0492} &\textbf{0.0660\rlap{\textsuperscript{**}}}      \\
                     & R@10                      & 0.0604 & \underline{0.0922}   & 0.0825 & {0.0907}                 & 0.0742  & {0.0662}   & 0.0228 & 0.0653 & 0.0745 & {0.0772} & \textbf{0.0975\rlap{\textsuperscript{**}}}     \\
                     & R@20                      & 0.0817 & 0.1253   & 0.1160 & {\underline{0.1264}}     & 0.1039  &  {0.0917}  & 0.0354 & 0.0940 & 0.1084 & {0.1107} & \textbf{0.1327\rlap{\textsuperscript{**}}}      \\
\midrule
\multirow{3}{*}{Sports}               & R@5      & 0.0209 & 0.0433   & 0.0355 &  {\underline{0.0440}}  & 0.0329  &  {0.0267}  & 0.0116 & 0.0307 & 0.0380 & {0.0397} &\textbf{0.0456\rlap{\textsuperscript{**}}}  \\
                     & R@10                      & 0.0317 & 0.0639   & 0.0562 &  {\underline{0.0645}}  & 0.0495  &  {0.0394}  & 0.0170 & 0.0472 &  {0.0581} & {0.0617} & \textbf{0.0674\rlap{\textsuperscript{**}}}    \\
                     & R@20                      & 0.0468 & 0.0904   & 0.0836 &  {0.0908}  & 0.0725  &  {0.0553}  & 0.0273 & 0.0728 & 0.0882 & \underline{{0.0929}}  & \textbf{0.0976\rlap{\textsuperscript{**}}}    \\   
\midrule
\multirow{3}{*}{Phone}               & R@5       & 0.0405 & 0.0723   & 0.0660 &  {\underline{0.0727}}  & 0.0571  &  {0.0451}  & 0.0206 & 0.0404 & 0.0602 & {0.0612} & \textbf{0.0756\rlap{\textsuperscript{**}}}  \\
                     & R@10                      & 0.0590 & \underline{0.1054}   & 0.0986 &  {0.1043}  & 0.0861  &  {0.0641}  & 0.0371 & 0.0623 & 0.0898 & {0.0918} &\textbf{0.1131\rlap{\textsuperscript{**}}}    \\
                     & R@20                      & 0.0855 & \underline{0.1482}   & 0.1418 &  {0.1473}  & 0.1228  &  {0.0899}  & 0.0600 & 0.0939 &  {0.1293} &  {0.1353} & \textbf{0.1590\rlap{\textsuperscript{**}}}    \\

\midrule
\multirow{3}{*}{{Recipe}}               & R@5       & 0.0128 & 0.0204   & 0.0121 &  {\underline{0.0210}}  & 0.0182  &  {0.0172}  & 0.0157 & 0.0171 & 0.0181 &{0.0189} & \textbf{0.0219\rlap{\textsuperscript{**}}}  \\
                     & R@10                      & 0.0229 & {0.0320}   & 0.0212 &  \underline{0.0322}  & 0.0309  &  {0.0269}  & 0.0274 & 0.0295 & 0.0316 & {0.0313} & \textbf{0.0334\rlap{\textsuperscript{**}}}    \\
                     & R@20                      & 0.0371 & 0.0487   & 0.0343 &  {0.0490}  & 0.0499  &  {0.0412}  & 0.0443 & 0.0475 & \underline{0.0504} &{0.0493} & \textbf{0.0528\rlap{\textsuperscript{**}}}    \\

\bottomrule
\end{tabular}
}
% \vspace{0.1cm}
\end{table*}

%% file: table/merge.tex
\begin{table*}[t]
\centering
% \small
\setlength\tabcolsep{4.5pt}
\vspace{0.2cm}
\caption{Effect of multi-task learning and different types of GIDs. The best scores are marked in bold.}
% \vspace{-0.1cm}
\label{tab:abl-id}
\begin{tabular}{lcccccccc}
\toprule
& \multicolumn{2}{c}{\textbf{Beauty}}         & \multicolumn{2}{c}{\textbf{Sports}}         & \multicolumn{2}{c}{\textbf{Phone}}    & \multicolumn{2}{c}{\textbf{Recipe}}      \\ 
\cmidrule(lr){2-3} \cmidrule(lr){4-5} \cmidrule(lr){6-7} \cmidrule(lr){8-9}
& \textbf{Recall@5}  & \textbf{NDCG@5} & \textbf{Recall@5}  & \textbf{NDCG@5} & \textbf{Recall@5}  & \textbf{NDCG@5} & \textbf{Recall@5}  & \textbf{NDCG@5} \\
\midrule
ColaRec      & \textbf{0.0667}    & \textbf{0.0449}    
             &\textbf{0.0442}     & \textbf{0.0294}      
             & \textbf{0.0745}    & \textbf{0.0490}    
             & \textbf{0.0198}    & \textbf{0.0151} \\
\midrule
(1) w/o textual content & 0.0527   & 0.0346  
                        & 0.0364   & 0.0239            
                        & 0.0636   & 0.0426 
                        & 0.0181   & 0.0141 \\
(2) w/o indexing        & 0.0637   & 0.0428             
                        & 0.0422   & 0.0278             
                        & 0.0728   & {0.0487} 
                        & 0.0179   & 0.0142 \\
(3) w/o $\mathcal{L}_{bpr}$        & 0.0612            & 0.0412           
                                   &{0.0424} &{0.0282} 
                                   & 0.0719            & 0.0486 
                                   & 0.0184            & 0.0140 \\
(4) w/o $\mathcal{L}_{c}$          &{0.0657} &{0.0434} 
                                   & 0.0422            & 0.0279             
                                   &{0.0731} & 0.0485 
                                   &{0.0188} & {0.0145} \\
\bottomrule
(1) $iad$    & 0.0658             & 0.0437             
             & {0.0428} & {0.0285}  
             & {0.0719} & 0.0474  
             & 0.0189             & 0.0145 \\
(2) $Random$ & 0.0600             & 0.0401             
             & 0.0411             & 0.0272             
             & 0.0667             & 0.0443   
             & {0.0190}             & {0.0149} \\
(3) $Content$ & {0.0662} & {0.0440} 
              & 0.0423             & 0.0278             
              & 0.0716             & {0.0477} 
              & 0.0183             & 0.0141 \\
\bottomrule

\end{tabular}
% \vspace{0.1cm}
\end{table*}

%% file: pics/part-length.tex
\begin{figure}
	\centering
	\subfigure[Beauty]{
		\begin{minipage}[t]{0.49\linewidth}
			\centering
			\includegraphics[width=1.7in]{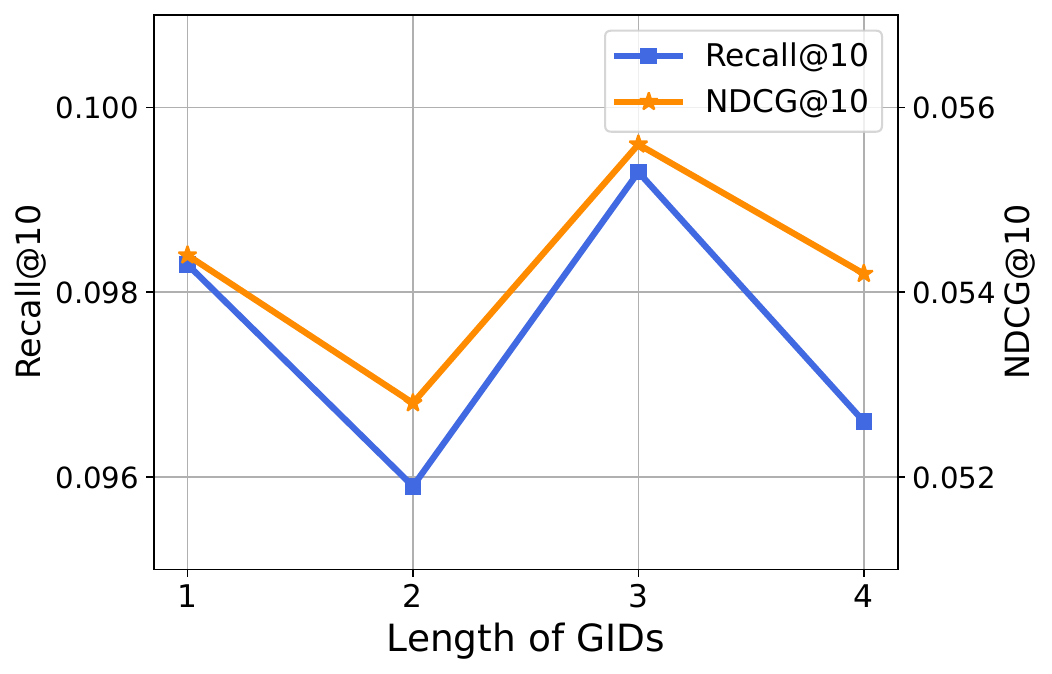}
		\end{minipage}
	}%
	\subfigure[Sports]{
		\begin{minipage}[t]{0.49\linewidth}
			\centering
			\includegraphics[width=1.7in]{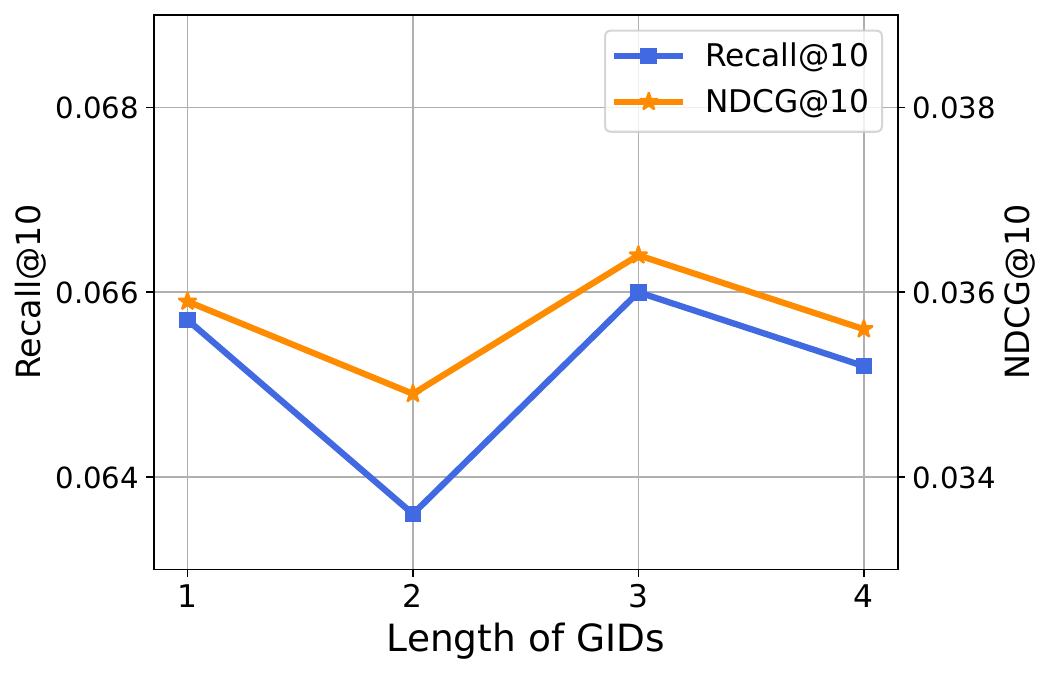}
		\end{minipage}
	}%
	%\caption{Analysis of the length of GIDs. We fix other things and vary GID length $l$ from $1$ to $4$.}
    \caption{Impact of the length of GIDs.}
	\label{fig:length}
 % \vspace{0.1cm}
\end{figure}

%% file: pics/part-cluster.tex
% 聚类个数K
\begin{figure}
	\centering
	\subfigure[Beauty]{
		\begin{minipage}[t]{0.49\linewidth}
			\centering
			\includegraphics[width=1.7in]{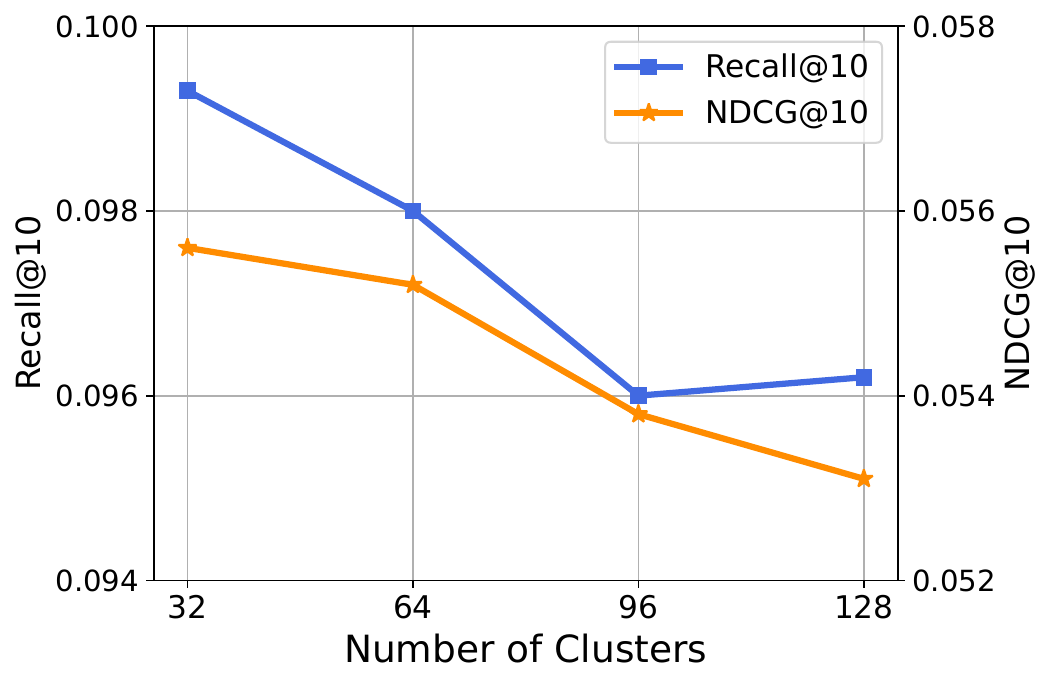}
		\end{minipage}
	}%
	\subfigure[Sports]{
		\begin{minipage}[t]{0.49\linewidth}
			\centering
			\includegraphics[width=1.7in]{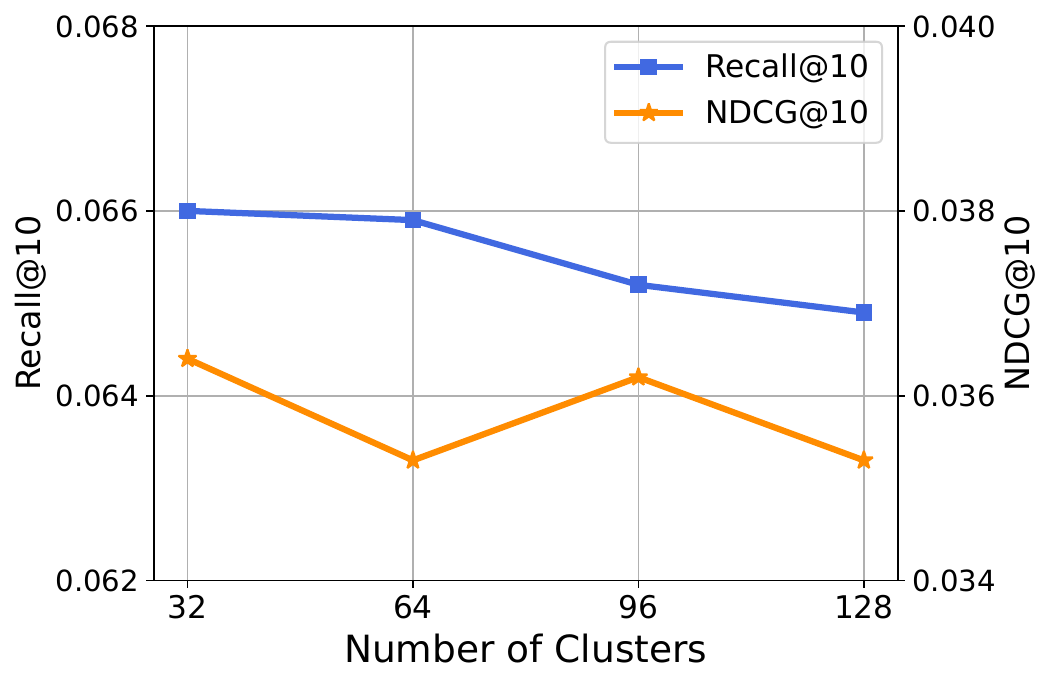}
		\end{minipage}
	}%
	\caption{Impact of the number of clusters. }
	\label{fig:center}
 % \vspace{0.1cm}
\end{figure}

%% file: sections/conclusion.tex
\section{Conclusions and future work}
This paper proposes ColaRec, a novel framework to conduct content-based collaborative generation for recommender systems. 
As an end-to-end generative recommender, ColaRec effectively integrates both item content information and user-item collaborative signals within a unified framework through a generative model.
In addition, an auxiliary item indexing task and a contrastive loss are proposed to further align the model's representations of item content and user-item collaborative signals.
We have conducted extensive experiments on four real-world datasets and empirical results demonstrate the effectiveness of ColaRec.

In the future, we intend to investigate more methods to construct  generative identifiers (GIDs) and adopt more effective approaches to better align content information and collaborative signals. 
% We also plan to introduce larger language models together with larger volume of training data to generate better recommendation. 
Negative sampling for generative recommendation is also one of our future works, either using GIDs to sample more informative negative samples or utilizing generative models to generate synthetic negative instances.
Besides, how to improve model efficiency for generative recommendation is also a potential research direction.

%% file: sections/acknowledge.tex
\section*{Acknowledgements}
This work was supported by the Tencent WeChat Rhino-Bird Focused Research Program (WXG-FR-2023-07), the Natural Science Foundation of China (62202271, 62372275, 62272274, T2293773,\\ 
62102234, 62072279), the National Key R\&D Program of China with grant No.2022YFC3303004, the Natural Science Foundation of Shandong Province (ZR2021QF129) and the Young Elite Scientists Sponsorship Program by CAST (2023QNRC001).  All content represents the opinion of the authors, which is not necessarily shared or endorsed by their respective employers and/or sponsors.